\documentclass{ws-procs9x6}

\newcommand{\xbj}{x}
\newcommand{\sst}{|\vec{S}_T|}

\begin{document}

\title{Single-spin asymmetries and Qiu-Sterman effect(s)
        \footnote{\uppercase{T}his work is supported by the \uppercase{A}lexander von
    \uppercase{H}umboldt foundation.}}

\author{A.\ Bacchetta}

\address{Theoretische Physik, Universit\"at Regensburg \\ 
D-93040 Regensburg, Germany} 

\address{Theory Group, Deutsches Elektronen-Synchroton DESY, \\
D-22603 Hamburg, Germany \\
E-mail: alessandro.bacchetta@desy.de}

\maketitle

\abstracts{
I discuss the relation between the Qiu-Sterman effects on one hand and the
Collins, Sivers and Boer-Mulders effects on the other hand. 
It was suggested before that 
some of these effects are in fact the same, thus providing interesting
connections between transverse-momentum dependent twist-2 functions and
collinear twist-3 functions. Here I propose an alternative way to reach similar conclusions.
}

\section{Introduction}
\label{intro}

Single-spin asymmetries have been observed in semi-inclusive deep inelastic
scattering and proton-proton collisions.\cite{Airapetian:2004tw,Adams:2003fx} 
Seemingly different mechanisms have
been advocated to explain these effects.
Qiu and Sterman proposed three possibilities, which I shall call chiral-even
distribution, chiral-odd distribution and chiral-odd fragmentation Qiu-Sterman
effects.\cite{Qiu:1991pp} Earlier work in the same direction was carried out
by Efremov and Teryaev.\cite{Efremov:1984ip} 
On the other hand, the Sivers\cite{Sivers:1990cc}, Boer-Mulders
\cite{Boer:1997nt} and Collins\cite{Collins:1993kk} effects can give rise to
the same asymmetries. 
It was argued by Boer, Mulders and Pijlman\cite{Boer:2003cm}, that
these mechanisms are related, as all of them involve gluonic-pole matrix
elements. 
This conclusion is apparently surprising, since the Sivers, Boer-Mulders and
Collins effects
can be described by {\em T-odd}, twist-2
distribution or fragmentation functions
depending
on intrinsic transverse momentum, while the effects discussed by Qiu-Sterman
are T-odd, twist-3, and collinear.
In my talk, I shall present an alternative derivation of the connection
between the Sivers function and the Qiu-Sterman chiral-even 
distribution functions. A similar relation should hold also 
between the Boer-Mulders
function and the Qiu-Sterman chiral-odd distribution function. 
On the other hand, I shall argue that the Collins function has a
different origin compared to the Qiu-Sterman chiral-odd fragmentation function.

\section{Semi-inclusive deep inelastic scattering}

Single-spin asymmetries in deep inelastic scattering have been discussed in a
large number of papers. I will now focus on the single-spin asymmetry that can
be observed in the process $l p^{\uparrow}\to l\pi X$ 
when the target is transversely polarized and the transverse
momentum of the final hadron is integrated over.\footnote{A similar case is when the 
final-state lepton is integrated over, but the transverse momentum of the
hadron is detected.}
This particular asymmetry has
been studied in a few references~\cite{Jaffe:1993xb,Boer:1997nt,Anselmino:1999gd} in terms of T-odd
distribution or fragmentation functions, while it has been studied in
by Koike\cite{Koike:2002ti} in terms of Qiu-Sterman effects. No experimental measurement has been attempted so
far, but it should be feasible at HERMES and COMPASS.

The general formula for the asymmetry up to subleading twist
is~\cite{Boer:1997nt}
\begin{equation}
A_{UT} =   
\frac{\sst}{d \sigma_{UU}}\sum_{q} \frac{2 \alpha^2 e_q^2}{s \xbj y^2}\, V(y)
\sin{\phi_S}\,\frac{M}{Q}\biggl[\xbj\, f_T^q  D_1^q
   - \frac{M_h}{M} \, h_{1}^{q} \frac{\tilde{H}^{q}}{z}\biggr],
\label{AUT1}
\end{equation} 
where 
$V(y) =  2\,(2-y)\, \sqrt{1-y}$.
 
The function $f_T$ is a twist-3 distribution function and 
can be split in two parts, an interaction-dependent part, which can be related
via equation of motions to quark-gluon-quark correlations, and a
Wandzura-Wilczek part, which is related to a twist-2
distribution function, in this case the Sivers function:\cite{Boer:1997bw}
\begin{equation}  
\xbj\, f_T = \xbj\,\tilde{f}_T^q - f_{1T}^{\perp (1) q}. 
\end{equation} 

Eq.~(\ref{AUT1}) becomes
\begin{equation}
A_{UT} =   
\frac{\sst}{d \sigma_{UU}}\sum_{q} \frac{2 \alpha^2 e_q^2}{s \xbj y^2}\, V(y)
\sin{\phi_S}\,\frac{M}{Q}\biggl[\bigl(\xbj\,\tilde{f}_T^q - f_{1T}^{\perp (1) q}\bigr)  D_1^q
   - \frac{M_h}{M} \, h_{1}^{q} \frac{\tilde{H}^{q}}{z}\biggr],
\label{AUT}
\end{equation} 

Eq.\ (\ref{AUT}) contains two different 
kinds of twist-3 contributions. As mentioned before,
the terms with a tilde are related to quark-gluon-quark correlations. They
don't vanish even if transverse momentum is naively neglected.
They
can be called {\em dynamical} twist-3 terms and should be related to the
Qiu-Sterman contributions studied by Koike\cite{Koike:2002ti} (specifically to the
chiral-even distribution $G$ and chiral-odd fragmentation $\hat{E}$ described
in his work).  The
term $f_{1T}^{\perp (1)}$ denotes the first moment (in transverse momentum
space) of the Sivers function. This term would vanish if a collinear
approximation was adopted from the beginning. 
Dynamically, it is a twist-2 term coming from
the gauge-link contribution to quark-quark correlations, but it
appears at twist 3 due to the fact that in this particular asymmetry (and in
general whenever the
transverse momentum of the outgoing hadron is not observed) off-collinear 
effects are kinematically suppressed. 
It can be therefore called a
{\em kinematical} twist-3 term and it has been studied by Anselmino {\it et
  al.}.\cite{Anselmino:1999gd}   
Note that in this asymmetry there is
no contribution involving the Collins function
due to off-collinear effects.

I come now to the main point of my talk. The $A_{UT}$ asymmetry
of Eq.~(\ref{AUT}) can be
calculated also for totally inclusive deep-inelastic scattering (by replacing
$D_1(z) \to \delta(1-z)$, $\tilde{H} \to 0$) and reduces to
\begin{equation}
A_{UT} =   
\frac{\sst}{d \sigma_{UU}}\sum_{q} \frac{2 \alpha^2 e_q^2}{s \xbj y^2}\, V(y)
\sin{\phi_S}\,\frac{M}{Q}\bigl(\xbj\,\tilde{f}_T^q - f_{1T}^{\perp (1)
  q}\bigr), 
\label{AUT3}
\end{equation} 
However, in totally inclusive deep-inelastic scattering
time-reversal invariance forbids the presence of such an asymmetry.\cite{Anselmino:1995gn,Jaffe:1989xx}
A relation is implied by this observation, namely
\begin{equation}  
\xbj\,\tilde{f}_T^q(x) - f_{1T}^{\perp (1) q}(x)  = 0, 
\label{main}
\end{equation} 
In principle, the relation holds only for the sum over all quark
flavors, but repeating the above argument for a hypothetical photon that 
couples selectively to different flavors, one obtains the above relation.

Eq.\ (\ref{main}) is the main result presented in this talk and provides a
complementary way to state that there is a relation between the chiral-even 
Qiu-Sterman distribution function and the first moment of the 
Sivers function. Note that the vanishing of the function $f_T$, 
which is equivalent 
to Eq.\ (\ref{main}), was already discussed by Goeke, Metz and 
Schlegel.\cite{Goeke:2005hb}  

An appropriate treatment of the T-odd distribution functions up to twist-3
should lead to the same result from a more formal point of view, making
clear that $\xbj\,\tilde{f}_T$ and $f_{1T}^{\perp (1)}$ are indeed the same
object and both originate from gluonic-pole matrix elements.\cite{Boer:2003cm}

%At this point, I want to stress a difference between my conclusions and those
%suggested by Teryaev in his talk at this workshop.\cite{Teryaev:2005} 
%We agree that there is
%a relation between the Sivers function and the Qiu-Sterman
%chiral-even distribution function. However, my point of view is that {\em
%  both} of them have to be taken into account separately and they eventually
%cancel in the asymmetry, while Teryaev considers this to be a double
%counting. In particular, I expect no $(\sin \phi_S)$ Fourier contribution to 
%the $A_{UT}$ asymmetry in
%deep-inelastic photoproduction, due to the absence of the chiral-odd
%fragmentation function $\tilde{H}$. 

Note that the asymmetry in Eq.\ (\ref{AUT}) -- once the first term is dropped --
turns out to be a good way to measure transversity, in particular in
experiments which are sensitive to higher twist
observables.\cite{Mulders:1996dh}
The function $\tilde{H}$ was introduced for the first time by Jaffe
and Ji,\cite{Jaffe:1993xb} who called it $\hat{e}_{\bar{1}}$. 
The absence of the Collins function in Eq.~(\ref{AUT}) suggests that it is
intrinsically different from $\tilde{H}$. In fact, in the literature it was
already observed that the Collins function is not related to gluonic
poles.\cite{Collins:2004nx,Amrath:2005gv}

\section{Drell-Yan}

In analogy to semi-inclusive DIS, we can consider $A_T$ asymmetries in
Drell-Yan processes, $p p^{\uparrow}\to l \bar{l} X$, integrated over the
transverse momentum of the lepton pair. 
This asymmetry has been
discussed by Boer, Mulders and Teryaev,\cite{Boer:1997bw} but the conclusions
reached by those authors are incomplete due to the fact that at that time the 
 gauge link was not taken into account as a source of T-odd effects.
The only contributions to the 
$A_T$ asymmetry should be (assuming proton $A$ to be
transversely polarized)
\begin{multline} 
A_T \propto \frac{\sst}{d \sigma_{UU}} 
\sum_{q} e_q^2\, 
\sin{\phi_S}\,\frac{M}{Q}\biggl[\xbj_{\scriptscriptstyle{A}} \Bigl((1-c)\, f_T^q +c\,\tilde{f}_T^q\Bigr)\,  f_1^{\bar{q}}
\\
   -  h_{1}^{q}\, \xbj_{\scriptscriptstyle{B}}\Bigl(c\, h^{\bar{q}} +(1-c)\,\tilde{h}^{\bar{q}} \Bigr)\biggr],
\label{AT}
\end{multline} 
where the factor $c$ depends on the frame of reference that is used to define
the azimuthal angle $\phi_S$ and can assume values between 0 and
1.\cite{Boer:1997bw} 

The first term of the asymmetry vanishes due to Eq.\ (\ref{main}). 
Through a formal treatment of twist-3 distribution functions,
it should be possible to prove that also
the function $h$ vanishes,\cite{Goeke:2005hb} implying a relation between the
Boer-Mulders function\cite{Boer:1997nt}
and the Qiu-Sterman chiral-odd distribution
function\cite{Kanazawa:2000kp} 
similar to Eq.\ (\ref{main}). The asymmetry reduces then to
\begin{equation} 
A_T \propto \frac{\sst}{d \sigma_{UU}} 
\sum_{q} e_q^2\, 
\sin{\phi_S}\,\frac{M}{Q}\biggl[c\,x_{\scriptscriptstyle{A}} \tilde{f}_T^q (x_{\scriptscriptstyle{A}})  f_1^{\bar{q}}(x_{\scriptscriptstyle{B}})
   - (1-c)\, h_{1}^{q} (x_{\scriptscriptstyle{A}}) x_{\scriptscriptstyle{B}}\,\tilde{h}^{\bar{q}} (x_{\scriptscriptstyle{B}})\biggr].
\label{AT2}
\end{equation} 
Note that, when defined in the frame of reference where $c=0$, this asymmetry 
gives the
opportunity to measure the transversity distribution function in
singly-polarized Drell-Yan, while in the frame where $c=1$ gives an
opportunity to study the Sivers function.

\section{Proton-proton collisions}

I now turn the attention to the $A_N$ asymmetry in the process $p p^{\uparrow}
\to \pi X$. The situation here is more involved than in 
deep inelastic scattering, due to the fact that partonic kinematics cannot
be reconstructed completely, in particular in the transverse
plane. Off-collinear kinematics at the partonic level 
has been analyzed in great detail by
Anselmino {\it et al.}.\cite{Anselmino:2005sh} It turns out that 
several T-odd distribution and
fragmentation functions can contribute to the $A_N$ asymmetry. These are again
kinematical twist-3 contributions, in the sense that they involve twist-2
functions with a kinematical suppression due to off-collinear kinematics.
Dynamical twist-3 effects in collinear kinematics are precisely those 
studied by Qiu and
Sterman.\cite{Qiu:1998ia}

As mentioned before, for distribution functions 
the two effects are identical and related to gluonic
poles. The partonic cross sections to be used in both cases should not be
normal partonic cross sections, but rather {\em gluonic-pole cross
sections}.    
An example of the use of gluonic-pole cross sections with the Sivers and
Boer-Mulders functions
has been given for 
the process $p p^{\uparrow} \to \pi \pi X$.\cite{Bacchetta:2005rm} 
Gluonic-pole cross sections are essentially equal to the standard partonic
cross sections multiplied by overall color factors.
Where do they come from and why they are not used in DIS and Drell-Yan? In
fact, they are already used in DIS and Drell-Yan, but they go somewhat
unnoticed! 
We know that T-odd functions arise
from gluonic poles present in the gauge link. In deep inelastic
scattering, the partonic process is $l q \to l q$. The gluons of the gauge
link can attach  {\em only} to the {\em outgoing} quark. The resulting
gluonic-pole cross section, $l \widehat{g q} \to l q$, 
in this simple case corresponds to the normal
partonic cross section. 
In Drell-Yan, the partonic process is $\bar{q}q \to l \bar{l}$, the gluon
can attach {\em only} to the {\em incoming} antiquark and the resulting
gluonic-pole cross section, $\bar{q} \widehat{g q} \to l \bar{l}$  is equal 
to {\em minus} the standard $\bar{q}q 
\to l \bar{l}$ cross section. 

In the partonic processes involved in $p p \to \pi X$,
colored partons are present both in the initial and the final state. The
resulting gluonic-pole cross sections are then equal to the standard partonic
cross section multiplied by nontrivial overall color factors, 
to be computed for each
individual process (and each individual channel of the process). 
Note that 
gluonic-pole cross sections have been studied only for the exchange of a
single gluon. It is not clear what happens when multiple gluon interactions
are taken into account.

For fragmentation functions the situation is different. Since the Collins
function is not related to gluonic poles, standard partonic cross sections can
be used with it, as done by Anselmino {\em et al.}.\cite{Anselmino:2004ky} 
On the contrary, gluonic-pole cross sections should be used
with the chiral-odd fragmentation function $\tilde{H}$.

\section{Conclusions}

I discussed the Qiu-Sterman effects on one hand and the
Sivers, Boer-Mulders and Collins functions on the other hand. I proposed a
relation between the chiral-even Qiu-Sterman distribution function and the
first moment of the Sivers function. A similar relation probably holds also 
between the Boer-Mulders function and the chiral-odd 
Qiu-Sterman distribution function.
On the contrary, I argued that the Qiu-Sterman chiral-odd
fragmentation function has a different origin compared to the Collins
function.

\section*{Acknowledgments}
I wish to thank D.\ Boer, M.\ Diehl, A.\ Metz, P.\ Mulders, F.\ Pijlman, O.\
Teryaev for fruitful
discussions. 
I thank the organizers for the invitation to this pleasant and interesting
workshop and ``Dipartimento
di Scienze e Tecnologie, Universit\`a del
Piemonte Orientale'' for financial support.

%%%%%%%%%%%%%%%%%%%%%%%%%%%%%%%%%%%%%%%%%%%%%%%%%%%%%%%%%%%%%%%%%%%%%%%%%%%
\bibliographystyle{h-physrev4}
\bibliography{mybiblio}

\begin{thebibliography}{10}

\bibitem{Airapetian:2004tw}
HERMES, A.~Airapetian {\em et~al.},
\newblock Phys. Rev. Lett. {\bf 94}, 012002 (2005), [hep-ex/0408013].
%%CITATION = HEP-EX 0408013;%%

\bibitem{Adams:2003fx}
STAR, J.~Adams {\em et~al.},
\newblock Phys. Rev. Lett. {\bf 92}, 171801 (2004), [hep-ex/0310058].
%%CITATION = HEP-EX 0310058;%%

\bibitem{Qiu:1991pp}
J.~Qiu and G.~Sterman,
\newblock Phys. Rev. Lett. {\bf 67}, 2264 (1991).
%%CITATION = PRLTA,67,2264;%%

\bibitem{Efremov:1984ip}
A.~V. Efremov and O.~V. Teryaev,
\newblock Phys. Lett. {\bf B150}, 383 (1985).
%%CITATION = PHLTA,B150,383;%%

\bibitem{Sivers:1990cc}
D.~W. Sivers,
\newblock Phys. Rev. {\bf D41}, 83 (1990).
%%CITATION = PHRVA,D41,83;%%

\bibitem{Boer:1997nt}
D.~Boer and P.~J. Mulders,
\newblock Phys. Rev. {\bf D57}, 5780 (1998), [hep-ph/9711485].
%%CITATION = HEP-PH 9711485;%%

\bibitem{Collins:1993kk}
J.~C. Collins,
\newblock Nucl. Phys. {\bf B396}, 161 (1993), [hep-ph/9208213].
%%CITATION = HEP-PH 9208213;%%

\bibitem{Boer:2003cm}
D.~Boer, P.~J. Mulders and F.~Pijlman,
\newblock Nucl. Phys. {\bf B667}, 201 (2003), [hep-ph/0303034].
%%CITATION = HEP-PH 0303034;%%

\bibitem{Jaffe:1993xb}
R.~L. Jaffe and X.~Ji,
\newblock Phys. Rev. Lett. {\bf 71}, 2547 (1993), [hep-ph/9307329].
%%CITATION = HEP-PH 9307329;%%

\bibitem{Anselmino:1999gd}
M.~Anselmino, M.~Boglione, J.~Hansson and F.~Murgia,
\newblock Eur. Phys. J. {\bf C13}, 519 (2000), [hep-ph/9906418].
%%CITATION = HEP-PH 9906418;%%

\bibitem{Koike:2002ti}
Y.~Koike,
\newblock AIP Conf. Proc. {\bf 675}, 449 (2003), [hep-ph/0210396].
%%CITATION = HEP-PH 0210396;%%

\bibitem{Boer:1997bw}
D.~Boer, P.~J. Mulders and O.~V. Teryaev,
\newblock Phys. Rev. {\bf D57}, 3057 (1998), [hep-ph/9710223].
%%CITATION = HEP-PH 9710223;%%

\bibitem{Anselmino:1995gn}
M.~Anselmino, A.~Efremov and E.~Leader,
\newblock Phys. Rept. {\bf 261}, 1 (1995), [hep-ph/9501369].
%%CITATION = HEP-PH 9501369;%%

\bibitem{Jaffe:1989xx}
R.~L. Jaffe,
\newblock Comments Nucl. Part. Phys. {\bf 19}, 239 (1990).
%%CITATION = CNPPA,19,239;%%

\bibitem{Goeke:2005hb}
K.~Goeke, A.~Metz and M.~Schlegel,
\newblock  Phys. Lett. B {\bf 618}, 90 (2005), [hep-ph/0504130].
%%CITATION = HEP-PH 0504130;%%

\bibitem{Mulders:1996dh}
P.~J. Mulders and R.~D. Tangerman,
\newblock Nucl. Phys. {\bf B461}, 197 (1996), [hep-ph/9510301],
\newblock Erratum-ibid.\ {\bf B484} (1996) 538.
%%CITATION = HEP-PH 9510301;%%

\bibitem{Collins:2004nx}
J.~C. Collins and A.~Metz,
\newblock Phys. Rev. Lett. {\bf 93}, 252001 (2004), [hep-ph/0408249].
%%CITATION = HEP-PH 0408249;%%

\bibitem{Amrath:2005gv}
D.~Amrath, A.~Bacchetta and A.~Metz,
\newblock Phys. Rev. {\bf D71}, 114018 (2005), [hep-ph/0504124].
%%CITATION = HEP-PH 0504124;%%

\bibitem{Kanazawa:2000kp}
Y.~Kanazawa and Y.~Koike,
\newblock Phys. Lett. {\bf B490}, 99 (2000), [hep-ph/0007272].
%%CITATION = HEP-PH 0007272;%%

\bibitem{Anselmino:2005sh}
M.~Anselmino {\em et~al.},
\newblock hep-ph/0509035.
%%CITATION = HEP-PH 0509035;%%

\bibitem{Qiu:1998ia}
J.~Qiu and G.~Sterman,
\newblock Phys. Rev. {\bf D59}, 014004 (1999), [hep-ph/9806356].
%%CITATION = HEP-PH 9806356;%%

\bibitem{Bacchetta:2005rm}
A.~Bacchetta, C.~J. Bomhof, P.~J. Mulders and F.~Pijlman,
\newblock Phys. Rev. {\bf D72}, 034030 (2005), [hep-ph/0505268].
%%CITATION = HEP-PH 0505268;%%

\bibitem{Anselmino:2004ky}
M.~Anselmino, M.~Boglione, U.~D'Alesio, E.~Leader and F.~Murgia,
\newblock Phys. Rev. {\bf D71}, 014002 (2005), [hep-ph/0408356].
%%CITATION = HEP-PH 0408356;%%

\end{thebibliography}
%%%%%%%%%%%%%%%%%%%%%%%%%%%%%%%%%%%%%%%%%%%%%%%%%%%%%%%%%%%%%%%%%%%%%%%%%%%%%

\end{document}